\documentclass[prb,aps,twocolumn,epsfig,eqsecnum]{revtex4}
\usepackage{epsfig}
\usepackage{wasysym}
\usepackage{bm}
\usepackage{pstricks}
\newlength{\myL}

\newcommand{\beq}{\begin{equation}}
\newcommand{\eeq}{\end{equation}}
\newcommand{\bea}{\begin{eqnarray}}
\newcommand{\eea}{\end{eqnarray}}

\newcommand{\sab}{S_{\alpha\beta}}
\newcommand{\dab}{\delta_{\alpha\beta}}
\newcommand{\mab}{M_{\alpha\beta}}
\newcommand{\nab}{N_{\alpha\beta}}
\newcommand{\fht}{\frac{h}{2}}

\newcommand{\tS}{{\tilde{S}}}
\newcommand{\tL}{{\tilde{L}}}

\newcommand{\tB}{{\tilde{B}}}
\newcommand{\thf}{{\tilde{h}}}
\newcommand{\tom}{{\tilde{\omega}}}

\newcommand{\etal}{{\em et al.}}

\newcommand{\Tr}{{\rm Tr}}

\def\tit#1#2#3#4#5{{#1}{\bf #2}, #3 (#4)}

\def\prl{Phys.\ Rev.\ Lett.\ }

\def\prb{Phys.\ Rev.\ B\ }

\def\jpco{J.\ Phys.\ Cond.\ Mat.\ }

\def\jpsj{J.\ Phys.\ Soc.\ Jpn.\ }

\def\cjp{Can.\ J. Phys.\ }

\begin{document}

\title{Semiclassical degeneracies and ordering 
for highly frustrated magnets in a field}

\author{S. R. Hassan$^{1,2,3}$ and R. Moessner$^3$}

\affiliation{$^1$Department de Physique,Universit\'e de Sherbrooke,
Qu\'ebec, Canada J1K 2R1}
\affiliation{$^2$CPHT, Ecole Polytechnique, 91128 Palaiseau Cedex, France}
\affiliation{$^3$Laboratoire de Physique Th\'eorique de l'Ecole Normale
Sup\'erieure, CNRS-UMR8549, Paris, France}

\begin{abstract}
We discuss ground state selection by quantum fluctuations in
frustrated magnets in a strong magnetic field. We show that there
exist dynamical symmetries -- one a generalisation of Henley's
gauge-like symmetry for collinear spins, the other the quantum relict
of non-collinear weathervane modes -- which ensure a partial survival
of the classical degeneracies. We illustrate these for the case of the
kagome magnet, where we find zero-point energy differences to be
rather small everywhere except near the collinear `up-up-down`
configurations, where there is rotational but not translational
symmetry breaking.  In the effective Hamiltonian, we demonstrate the
presence of a term sensitive to a topological `flux'. We discuss
the connection of such problems to gauge theories by casting the
frustrated lattices as medial lattices of appropriately chosen simplex
lattices, and in particular we show how the magnetic
field can be used to tune
the physical sector of the resulting gauge theories.
\end{abstract}


\maketitle

\section{Introduction}
The rich behaviour presented by geometrically frustrated magnets has
made them a popular subject of
study.\cite{schifferramirez} Particular
attention has been devoted to systems in which the spins reside on
corner-sharing units: the kagome lattice in $d=2$ dimensions,
consisting of corner-sharing triangles, and the $d=3$ pyrochlore
lattice, made up of corner-sharing tetrahedra, whose sparse
connectivity makes them particularly strongly frustrated.\cite{moecha}

Classical ground state degeneracies can be considered to be the
hallmark of such geometrical frustration. One natural question is what
happens to these ground state degeneracies in the presence of quantum
fluctuations. Generically, one would expect the degeneracies to be
lifted, so that they would show up in a quantum model as a large
density of states at low energies, on a scale set by the strength of
the quantum fluctuations. Indeed, such a large density of low-energy
excitations is perhaps the most striking feature of the kagome spin
$S=1/2$ Heisenberg magnet,\cite{schifferramirez} although as yet there
exists no agreement on its origin.

Given the difficulties in studying the limit of strong quantum
fluctuations, small $S$, it has proven to be fruitful to consider
instead the semiclassical limit,\cite{olegproc} $S\rightarrow\infty$,
where quantum fluctuations endow the classical ground states with an
effective energy functional via the zero-point energy of the spin
excitations, $\sum \hbar\tom(\{{\tS_i^{(0)}\}})/2$, where the $\tom$
denote the frequencies of the excitations around a ground state
configuration $\{{\tS_i^{(0)}}\}$. (In the following,
we set $\hbar=1$.)  Perhaps the most striking result was that a
sizeable, albeit reduced, degeneracy can survive in the semiclassical
limit. The most prominent example of this was again provided by the
kagome lattice, where the equations of motion for excitations around
{any} {\em coplanar} ground state can be written in an identical form,
so that all these states have the same zero-point
energy.\cite{chaholds} The next important step was Henley's
observation,\cite{henleyshort} sketched below, that degeneracies
for {\em collinear} ground states could be understood in terms of a
gauge-like symmetry in the equations of
motion.

In a related development, a great deal of attention has recently been
focused on the behaviour of frustrated magnets in an external field,
in particular on magnetisation plateaux and
magnetocalorics.\cite{zhitokagobdo,honecker,tsunezhito,penc,takagiplat,richtersat,cabrakagome,honeckerplateau}

In this paper, we follow up on those two strands of work by
considering the properties of frustrated magnets on corner-sharing
geometries in an external field in the semiclassical limit.  We start
from the observation that such corner-sharing lattices can be thought
of as medial lattices of a `simplex lattice'; the spins are bond
variables on the simplex lattice, whereas the corner-sharing units
reside on its sites.

The ground state condition imposed by classical Hamiltonian then takes
the form of a local constraint, which can conveniently encoded in a
gauge theory, the `physical' sector of which can be changed by the
application of a magnetic field, so that one can tune between, e.g.,
loop and dimer models in this fashion.

We then consider the harmonic theory of excitations in such magnets by
generalising the equations of motion derived in
Ref.~\onlinecite{moecha} to include a magnetic field. This will allow
us to show that Henley's gauge-like symmetry for collinear spins can
be generalised to this case, but in addition that a different
dynamical symmetry exists for non-collinear spin configurations.

We illustrate these ideas with the example of a Heisenberg magnet on
the kagome lattice in an external field. This choice allows us to
build on a substantial body of previous work (on classical equations
of motion, order by disorder and magnetisation plateaux,
\cite{chaholds,hbc,ritchey,shenhold,zhitokagobdo,cabrakagome,honeckerplateau,Chubukov}).In keeping with received
wisdom, the semiclassical order by disorder we find favours coplanar
or collinear states. In different field ranges, both states derived
from $Q=0$ and $Q=\sqrt{3}\times\sqrt{3}$ configurations occur; we
find $Z_3$ and $Z_3\times U(1)$ order parameters, some of which will
only be expressed at minuscule temperatures. For the value of the
magnetic field which permits a collinear configuration, we find that
the semiclassical ordering pattern differs from that proposed for
$S=1/2$.\cite{honeckerplateau}

We contrast this behaviour to the kagome lattice thought of as a
network of corner-sharing {\em hexagons}, where the effective
semiclassical model does not permit such simple solutions. Finally, we
comment on experiment, and close with some more general remarks about
gauge theories for this problem.

The remainder of this paper is organised as follows.  In
Sect.~\ref{sec:medlat}, we introduce notation by writing down the
Hamiltonian in a field for a simplex lattice and deriving
(Sect.~\ref{sec:eofmh}) the equations of motion. Sects.~\ref{sec:Dsym}
and~\ref{sec:Effectiveh} deal with the dynamical symmetries and the
properties of effective Hamiltonians.  Sect.~\ref{sec:Gsl} explores in
detail ground state selection in the kagome magnet, and
addresses  issues arising for the kagome-hexagon
model. We conclude in Sect.~\ref{sec:summary}.

\section{Medial lattice construction}
\label{sec:medlat}
A frustrated lattice made up of corner-sharing units (also known as
simplices), can be obtained by considering any lattice $\Lambda$, which
we will refer to as the simplex lattice. The midpoints of the bonds of
the simplex lattice define its medial lattice, $\Upsilon$.

\begin{figure}
  \includegraphics[width=\columnwidth]{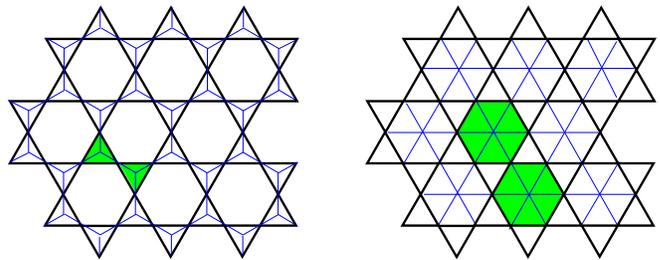} 
\caption{(colour online) The
  kagome lattice (fat lines) is the medial lattice of both the
  honeycomb and the triangular lattices (thin lines).  The former
  corresponds to a lattice of corner-sharing triangles, while the
  latter consists of corner-sharing hexagons.}
\label{fig:kagomemedial}
\end{figure}

The choices for $\Lambda$ of the square, honeycomb and diamond
lattices yield the well-known checkerboard, kagome and pyrochlore
lattices as their respective medial counterparts $\Upsilon$.  In fact,
there exists a second simplex lattice, the triangular lattice, which
has the kagome lattice as its medial lattice $\Upsilon$ (see
Fig.~\ref{fig:kagomemedial}). As discussed further down, systems on
medial lattices generally lend themselves to a gauge theoretic
description, for concrete instances see
Refs.~\onlinecite{misserpaskag,hermelepyro,sentmot}

Since sites $i\in\Upsilon$ are also links $\alpha\beta$ (with $\alpha,
\beta \in \Lambda$) on the simplex lattice, we can alternatively 
denote the spin $\tS_i$ by a link variable, $\tS_{\alpha\beta}$. A
brief remark on notation: we denote the location of a spin as
subscripts, its components as superscripts, and the spin length by $S$
(without super- or subscripts). 

The ground state condition of the frustrated spin Hamiltonian now
imposes constraints on the values of the link variables
$\tS_{\alpha\beta}$ emanating from a site $\alpha\in\Lambda$.  For
instance, denoting the set of links (i.e., sites of $\Upsilon$) by
$\upsilon(\alpha)$, the nearest neighbour exchange Hamiltonian is
given by
\bea
H_J=J\sum_{\langle ij\rangle}S_i \cdot S_j
=\frac{J}{2}\sum_\alpha {\tL}^2_\alpha\ ,
\label{eq:jexch}
\eea
where $\tL_\alpha=\sum_{i\in\upsilon(\alpha)}\tS_i$.  

The ground state condition $L_\alpha=0$ $\forall \alpha$ represents
such a local constraint. An external magnetic field provides a handle
for tuning this constraint. Adding a Zeeman term (where $g$ is the
g-factor, $\mu_B$ the Bohr magneton and $\tB$ the magnetic field),
\bea
H_Z=-\sum_i g \mu_B \tS_i\cdot \tB
\label{eq:zzee}
\eea
yields the Hamiltonian,
\bea
%\label{eq:hamijz}
\nonumber
H&=&H_J+H_Z=\frac{J}{2}\sum_\alpha 
({\tL}^2_\alpha-\frac{g\mu_B}{J}\tB\cdot \tL_\alpha)\\
&=& \frac{J}{2}\sum_\alpha ({\tL}_\alpha-\thf/2)^2\
=\frac{JS^2}{2}\sum_\alpha ({L}_\alpha-h/2)^2\ ,
%\label{eq:hamijz}
\nonumber\\
&\equiv& \frac{JS^2}{2}\sum_\alpha {l}_\alpha^2\
\label{eq:hamijz}
\eea
where $\thf=g \mu_B \tB /J$, and where we have dropped an
($\thf$-dependent) constant. All states with $l_\alpha\equiv
L_\alpha-h/2=0\ $ $\forall \alpha$ are thus ground states, provided
such states exist.  Here and in the following, expressions for spins,
fields and frequencies are rescaled by the spin length $S$ as in
$L_\alpha=\tL_\alpha/S$.

For simplicity, let us first consider the case of Ising spins ($n=1$,
$S=\pm1$) for which the $L_\alpha$ are integers between $\pm
z_\alpha$, where $z_\alpha$ is the coordination of site
$\alpha\in\Lambda$. Effectively collinear Ising spin configurations
can also arise in highly frustrated continuous spin models, via
different forms of order by
disorder.\cite{shenderobdo,henleycoll,moecha,penc,takagiplat,obdist} 
If $z$ is even
(odd), so are the possible $L_\alpha$.

If we denote spins pointing up by a link occupied by a dimer, and
spins pointing down by an empty link, one observes that $L=z$
corresponds to the non-degenerate saturated state with all spins
aligned, and all links empty. As has been noted in many contexts,
$L=z-2$ is equivalent to one link emanating from each site being
occupied. A magnet in a field on the simplex lattice $\Upsilon$ is
thus equivalent to the hardcore dimer models on its simplex lattice
$\Lambda$. Similarly, $L=z-4$ corresponds to loop models, which are
fully packed in the sense that each site is part of a loop. A familiar
example of this is the pyrochlore magnet in zero field.

\begin{figure}
  \includegraphics[width=\columnwidth]{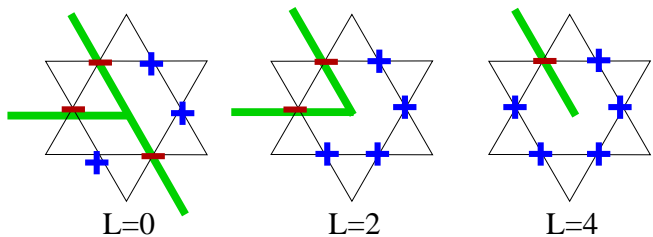}
  \caption{(colour online)
Multi-dimer configuration corresponding to different values of 
the hexagon magnetisation, $L$. Besides the usual dimer model below
saturation ($L=4$), the $L=2$ sector defines a loop model, and $L=0$ 
a three-dimer model.
}
\label{fig:kagomespace}
\end{figure}

For sufficiently large $z$, this continues for further $L\geq 0$. We
have illustrated these cases in Fig.~\ref{fig:kagomespace}, where
$L=0$ corresponds to the easy-axis kagome model of
Ref.~\onlinecite{BFG}, which from the current perspective is a three
dimer model on the triangular lattice.

For Ising spins, at the transition field where spin $L=z-2n$ becomes
degenerate with $L=z-2n+2$, the allowed vertex configurations are
those of both cases. This is in particular the case for odd $z$ in
zero field, and it goes along with a spike in the zero temperature
entropy at the field strength in question.

\section{Equations of motion}
\label{sec:eofmh}
The simple form of the Hamiltonian in terms of the $L$ suggests that
these variables should be useful quantities to use when constructing
effective theories. Indeed, the thermodynamics of the pyrochlore
Heisenberg antiferromagnet is well captured by formulating a theory in
terms of the $L$ of a single tetrahedron.\cite{moeberl}

Similarly, the $L$ are useful quantities when discussing the spectrum
of excitations around a classical ground state. In zero field, changes
of the spin orientation that keep $L\equiv0$ cost zero energy. Hence,
all finite frequency modes involve $L\neq0$. Indeed, the zero field
equations of motion for Heisenberg spins around a ground state
configuration $\{S^{(0)}\}$ is\cite{moecha}
\bea
\dot{L}_\alpha= 
JS\sum_{\beta:\alpha}S^{(0)}_{\alpha\beta}\times L_\beta \ . \ 
\eea
Here and in the following, the symbol $\times$ denotes a vector
product, and the
sum $\sum_{\beta:\alpha}$ runs over all
$S_{\beta\alpha}\in\upsilon(\alpha)$. In the following, we drop the
superscript 0 as, in our harmonic theory, the spins $S_{\alpha\beta}$
will always refer to a ground state configuration.

By the same token, the simple form of Eq.~\ref{eq:hamijz} suggests
formulating the equations of motion in terms of 
\bea
l_\alpha = L_\alpha-h/2 \ .
\eea
The resulting form of the equations of motion have a similar structure
(see App.~\ref{app:eofmz}):
\bea
(\frac{d}{dt}-\frac{1}{2} h\times)\,l_\alpha=
JS\sum_{\beta:\alpha}S_{\alpha\beta}\times l_\beta \ .
\label{eq:eofmz}
\eea

\section{Dynamical symmetries}
\label{sec:Dsym}
\subsection{Collinear ground states}
For collinear spins in zero field, we have $L=(L^1,L^2,0)$ and
$S_i=(0,0,S^3)$. It was pointed out by Henley that the form of the
equations of motion remains unchanged when making the Ising
($\zeta=\pm1$) gauge-like variable transformation
\bea
L_\alpha\rightarrow\zeta_\alpha L_\alpha
\label{eq:tranf1}
\eea
together with 
\bea
\sab\rightarrow\zeta_\alpha\sab\zeta_\beta\   \ ,
\label{eq:tranf2}
\eea
for any set of $\{\zeta\}$.

Thus, provided the new state, with the $\sab$ changed in sign, is
again a ground state of the Hamiltonian (\ref{eq:hamijz}), the
spectrum of its excitation energies is going to be identical to that
of the starting state.

It can be seen that this continues to be the case for the equations of
motion for the $\{l\}$ in the presence of a field, which can be
seen by writing them out explicitly:
\bea
(\zeta_\alpha\dot{l}^1_\alpha)+\fht (\zeta_\alpha l_\alpha^2)
&=&-(\zeta_\alpha\sab^3\zeta_\beta)(\zeta_\beta l_\beta^2)\nonumber\\
(\zeta_\alpha\dot{l}^2_\alpha)-\fht (\zeta_\alpha l_\alpha^1)&=&
(\zeta_\alpha\sab^3\zeta_\beta)(\zeta_\beta l_\beta^1) 
\label{eq:eofmznoncoll}\\
\dot{l}^3_\alpha&=&0\nonumber\ ,
\eea
where we take the field along the $3$-direction.

For a non-collinear state, this transformation does not work as the
bottom member of Eq.~\ref{eq:eofmznoncoll} does not hold in that case.
This implies that all spins need to point along the field axis, as in
the ground state $L_\alpha=h/2$ implies that not all
$S^3\in\upsilon(\alpha)$ can be zero.

Further down, we show how the kagome magnet at 1/3 of its saturation
field provides an example of such a symmetry, including a subextensive
degeneracy as well as a resulting $Z_3$ order parameter.

\subsection{Non-collinear ground states}
There exists a different dynamical symmetry for the case of
non-collinear spins. This can be seen by writing the equations of motion in
component form:
\bea
\dot{l}^1_\alpha+\fht l_\alpha^2&=&
(\eta\sab^2)(\eta l_\beta^3)-\sab^3l_\beta^2
\nonumber\\
\dot{l}^2_\alpha-\fht l_\alpha^1&=&
\sab^3l_\beta^1-(\eta\sab^1)(\eta l_\beta^3) 
\label{eq:eofmzcomp}\\
(\eta \dot{l}^3_\alpha)&=&(\eta\sab^1)l_\beta^2-(\eta\sab^2)l_\beta^1
\nonumber\ .
\eea
Eq.~\ref{eq:eofmz} corresponds to $\eta=1$. 

For $\eta=-1$, these equations retain their form under
the transformation
\bea
l_{\alpha}^3\rightarrow\eta l_\alpha^3\ \ ;\ 
\sab^{1,2}\rightarrow\eta\sab^{1,2} \ .
\label{eq:copgauge}
\eea
Notice that this transformation is distinct from the collinear gauge
transformation: for collinear states, $l^3\equiv S^{1,2}\equiv0$, and the
transformation does not generate a new spin state.

This transformation needs to be applied consistently: from the bottom
member of Eq.~\ref{eq:eofmzcomp} it follows that a change
(\ref{eq:copgauge}) involving the plaquette $\beta$ requires a concomitant
change on those neighbouring $\alpha$ plaquettes for which
$|\sab^3|\neq 1$. If all $|\sab^3|\neq1$, one therefore 
ends up transforming all the spins in the system, which amounts to a
global symmetry operation. 

However, if one can find a set of spins the neighbours of which all
point in the $3$-direction, the transformation is non-trivial. This is
the same condition as that for the weathervane modes of the classical
kagome magnet in zero field.\cite{ritchey} The local dynamical $Z_2$
symmetry in the semiclassical problem is therefore the vestige of the
`accidental' local $U(1)$ symmetry present in the classical case, and
it persists to the case of nonzero fields.

Here, we have demonstrated that it holds not only for the zero-field
case, in which the angle between any two neighbouring spins is
uniform.  Below we provide an example of this for the kagome magnet in
a field, and discuss the order parameter resulting from the ensemble
of semiclassically degenerate states.

\section{Effective Hamiltonians}
\label{sec:Effectiveh}
The basic idea of effective Hamiltonians is to discard as much
microscopic information as possible, essentially by
integrating out fluctuations, in the hope that this procedure will
give rise to an expression which is simple 
(and, ideally, local).\cite{henleyshort}

\subsection{Selection of collinear configurations}
It has been suggested, first in the seminal work by Shender on quantum
order by disorder, that quantum fluctuations favour collinear spin
configurations:\cite{shenderobdo,henleycoll} integrating them out in a
semiclassical framework leads, crudely, to an effective biquadratic
Hamiltonian of the form
\bea
H_4=J_4 S\sum_{\langle ij\rangle} (S_i\cdot S_j)^2\ .
\eea

We have tested this assertion in the present context by considering
the $Q=0$ configurations at a magnetic field where a collinear state
is possible: at $h=2$, the up-up-down state is possible, in which the
spins on two sublattices of the kagome lattice point up, and down on
the third (Fig.~\ref{fig:Q0string}).

\begin{figure}
\includegraphics[width=0.3\columnwidth]{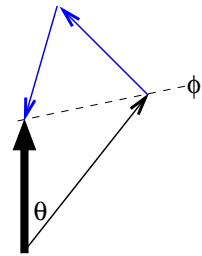} 
\caption{(colour online) Parametrisation 
of the $Q=0$ ground states. The fat arrow denotes the magnetic field;
its length is $h/2$. $\phi$ denotes the rotation angle of the top pair
of spins out of the plane.}
\label{fig:3spinh}
\end{figure}

\begin{figure}
 \includegraphics[width=\columnwidth]{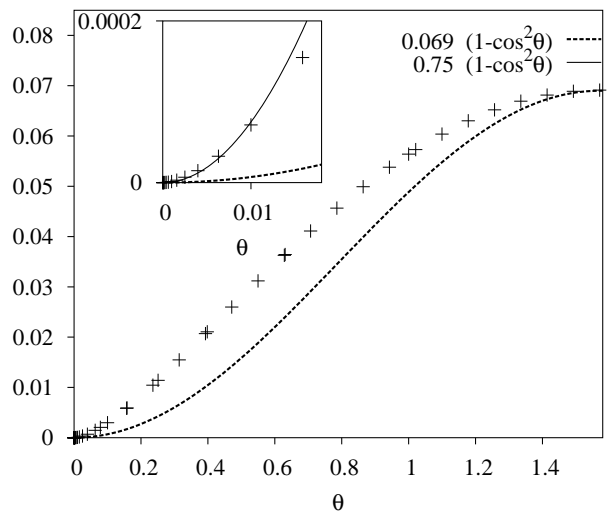} 
\caption{Zero-point energy per spin, in units of $JS$, at $h/2=1$ as a 
function of $\theta$ (Fig.~\ref{fig:3spinh}).}
\label{fig:eoftplat}
\end{figure}

At $h=2$, there is in fact a two-parameter family of possible $Q=0$
states, as depicted in Fig.~\ref{fig:3spinh}. In
Fig.~\ref{fig:eoftplat}, we have evaluated the zero-point energy as a
function of $\theta$, that is to say for coplanar spin
configurations. Here, $\theta=0$ corresponds to the collinear state.

From this picture, two conclusions can be drawn. Firstly, the
biquadratic form broadly has the correct shape, with the minimum and
maximum in the right places, and a monotonic behaviour in
between. Secondly, the shape of the curve is in quantitative
disagreement with the simple biquadratic term. Therefore, when trying
to extract the effective $J_4$, one should proceed in two different
ways, depending on what information one is interested in.

If the question one is asking concerns the overall energy scale at
which quantum fluctuations start to assert themselves, one is
basically interested in the difference between the highest and the
lowest point on that curve. By contrast, if one starts from the lowest
temperatures and asks about the energy cost of small deviations from
collinearity, one needs to fit the curvature at low $\theta$. (In
fact, fitting to a power law gives an exponent deviating substantially
from 2 already for small $\theta$; however, this exponent does seem to
extrapolate to 2 within the limits of our numerical accuracy).

Note that the values for $J_4$ thus obtained differ by an order of
magnitude when the low-$\theta$ fit is done to be accurate around
$\theta=5^\circ$ (inset of Fig.~\ref{fig:eoftplat}). This discrepancy
continues to grow as the range of $\theta$ is lowered. 

Is the form of the effective biquadratic interaction more accurate in
the absence of a field? For the kagome magnet, there are no collinear
states available in zero field. However, for the checkerboard (planar
pyrochlore) magnet, such configurations do exist, and we have checked
that the corresponding $E_0(\theta)$ has the same behaviour in that
case.

\subsection{Gauge form of Hamiltonian} 
By defining $l^\pm=l^1\pm i l^2$, one can rewrite the equations of
motion for collinear spins as
\bea
\pm \omega l^\pm_\alpha=(\fht\delta_{\alpha\beta}+ \sab)l_\beta^\pm\ .
\eea
To obtain $E_0=JS\langle|\omega|\rangle/2$, this form allows us to
make use of Henley's trick for calculating the zero point energy by
writing 
\bea
\Tr |\omega|&=&\Tr\sqrt{\omega^2}\\ \nonumber
&=&\Tr\sqrt{\left(\frac{h^2}{4}+z\right)\dab+h\sab+\mab}\ ,
\eea
and then Taylor expanding the square root to evaluate the trace term
by term. Here $z$ is the coordination of the simplex lattice, and
$\mab=S_{\alpha\gamma}S_{\gamma\beta}-z\dab$, so that its diagonal
entries vanish. Thus, formally (and up to a constant), for
$\nab=-(h\sab+\mab)/\sqrt{{h^2}/{4}+z}$,
\bea
\Tr |\omega|&=&\sqrt{\frac{h^2}{4}+z}\ \Tr\sqrt{\dab-\nab}\nonumber\\
&=&-\sqrt{\frac{h^2}{4}+z}\ \sum_{m=2}^\infty 
\frac{(2m-3)!!}{2^mm!} \Tr N^m \ .
\label{eq:exptr}
\eea
$\Tr N^m$, as in zero field, only contains terms of the type
$S_{\alpha\beta}S_{\beta\gamma}...S_{\omega\alpha}$, which can be
identified with closed loops
$\alpha\rightarrow\beta\rightarrow\gamma\rightarrow\cdots\rightarrow\omega\rightarrow\alpha$
on the lattice $\Lambda$ as each site of $\Lambda$ occurs an even
number of times.

Besides altering the coefficients of different loops, the inclusion of
the magnetic field has changed the structure of this expression by
including a term linear in $\sab$ in the Taylor expansion, thus
allowing for the possibility that $\Tr N^m$ contains terms with an odd
number of factors of $\sab$. For bipartite lattices $\Lambda$, such
loops are not possible, but they do exist for nonbipartite ones, where
the structure of the expansion for $h\neq0$ thus contains terms which
are absent in zero field.

In any case, these loops have the form of fluxes $\Phi$ -- the factors
\bea
\Phi=S_{\alpha\beta}S_{\beta\gamma}...S_{\omega\alpha}
\label{eq:fluxS}
\eea have the
appearance of line integrals of the Ising `gauge field' $S$.  Knowing
the form of the terms appearing in the Hamiltonian, one still needs to
fix their coefficients. This problem has recently been adressed in
some detail by Hizi and Henley.\cite{henleyshort} 

Here, we make some further remarks on the structure of the relevant
solutions for the case that the dominant terms in the effective
Hamiltonian are those involving the elementary plaquettes of the
simplex lattice $\Lambda$. By elementary plaquettes we mean the
plaquettes the tiling of which makes up $\Lambda$, for instance
a square in the case of the square lattice.
 
For a two-dimensional lattice $\Lambda$ which is a tiling of such
elementary plaquettes, fixing the flux through the elementary
plaquettes determines the fluxes on all contractible loops of the
system. If the pattern of elementary fluxes is the same in two
different states, this means that the contribution to the energy is
the same in both states even for longer loops.

This does not mean that the energy is uniquely determined by
specifying the elementary fluxes: the fluxes through 
non-contractible loops such as a
loop winding around a cylinder in the case of periodic boundary
conditions are not thus determined. Below, we present a numerical
determination, for the case of the kagome lattice, that two states
having locally the same flux but differing in the flux through a
non-contractible loop have an zero-point energy difference which
approaches a constant, $JS$, in the thermodynamic limit.

Therefore, even though the energy measured locally in the flux
expansion is identical everywhere, there is a nonvanishing nonlocal
contribution to the energy.

\section{Ground state selection in the kagome magnet}
\label{sec:Gsl}
In this section, we give examples of the dynamical
symmetries discussed above, and discuss the effective models and order
parameters which arise.

\subsection{Collinear states}
\begin{figure}
 \includegraphics[width=0.8\columnwidth]{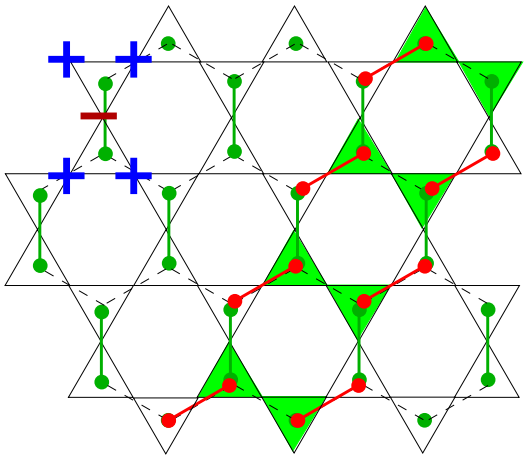} 
\caption{(colour online) The kagome magnet in a field $h=2$. 
Down spins are denoted by dimers. A gauge-like transformation is
carried out on a string of triangles: the spins at the corners of each
shaded triangle are flipped.}
\label{fig:Q0string}
\end{figure}

We start with the case of collinear spins at a magnetic field $h/2=1$,
where $L\equiv1$ allows for states in which two spins in each triangle
are aligned with the field, and one is antialigned.  As mentioned
above, these states map onto dimer model on honeycomb lattice, by
placing a dimer on links where $S_{\alpha\beta}$ points down
(Fig.~\ref{fig:Q0string}).

There are exponentially many such dimer coverings, leading to an
entropy of ${\cal S}=0.1077$ per spin once the collinear
configurations have been selected. The zero-point fluctuations will
lift the remaining degeneracy as different collinear configurations
have different `fluxes' through their elementary plaquettes. 

Two particularly simple representatives are the $Q=0$ configuration
(all dimers residing on vertical links, Fig.~\ref{fig:Q0string}), and
the $Q=\sqrt{3}\times\sqrt{3}$ configuration, which involves a
tripling of the unit cell. The former has uniformly zero flux in each
plaquette, whereas the latter has a flux of $\pi$ in two plaquettes
out of three. 

We have evaluated the zero-point energy for these states for a large
system (and for all collinear 
configurations of a finite system, see below); as
we disagree in part with several of the published results on this, we
have given details of the calculation in App.~\ref{app:eofmzr3r3}.
The fact that the $Q=0$ state has a lower zero would imply that
fluxless configurations are preferred.

It turns out to be easy to characterise these configurations
completely. Only the states in which exactly two dimers occur on each
hexagon are allowed, for periodic boundary conditions. For the readers
familiar with height models, these are the maximally tilted states, or
equivalently the non-flippable ones.

For a system of linear length $l$, with $3{l^2}$ spins and periodic
boundary conditions, we find the number $N_0$ of zero flux states to
be
\bea
N_{0}=3(2^l-1)\ .
\label{eq:Nnonflip}
\eea
However, not all of these can be related to one another via the
transformations described above.

To see this, let us consider in more detail what kind of
transformations are generated by (\ref{eq:tranf1}, \ref{eq:tranf2}).
Note that, applied to a single triangle $\alpha$, the total spins of
triangle $\alpha$ and the neighbouring $\beta$'s change, so that these
violate the ground-state condition. To undo this change, one needs to
apply the transformation to one-dimensional sets of triangles, as
depicted in Fig.~\ref{fig:Q0string}. This tranformation is analogous
to the one discussed by Tchernyshyov \etal\ for the planar pyrochlore
lattice; the zero-field case discussed there exhibits the additional
feature of independent transformations on different
sublattices.\cite{checkerboard}

The dimers of the resulting configuration, superimposed on the initial
configuration, form two `strings' (lines of alternating new and old
dimers). Even though a single string does not change the local fluxes,
such strings can only be created in pairs by the gauge transformation
(and they can be separated from one another by further transformations).

\begin{figure}
  \includegraphics[width=\columnwidth]{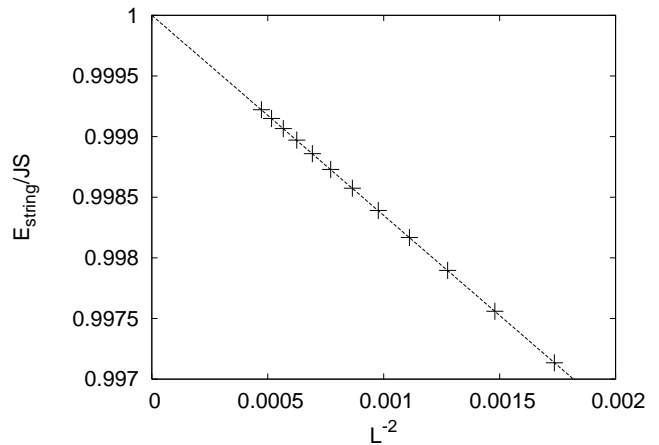} 
\caption{Energy cost of
  a topological `flux' in the collinear kagome ground state 
depcited in Fig.~\ref{fig:Q0string}, as a
  function of system size. }
\label{fig:topoQ0}
\end{figure}

One would therefore in general expect the configurations differing by
an odd number of strings to differ in energy; we have checked that
this is indeed the case, see Fig.~\ref{fig:topoQ0}: the energy cost
for an odd number of strings remarkably equals, to within our numerical
accuracy, precisely $JS$, which suggests that this is a robust result.
As the fluxes through any contractible loop
are the same independent of the parity of the number of strings, this
energy difference in the effective flux Hamiltonian is due to a
topological flux.

To find the number of ground states, we thus need to enumerate the
number of even, $N_0^g$, and odd, $N_0^u$ configurations
separately. For odd $l$, Eq.~\ref{eq:Nnonflip} gives the correct
number of ground states as the flux around a non-contractible loop
$\prod_\circ S_{\alpha\beta}$ always equals $-1$ in one direction. For
$l$ even, we find
\bea
N_0^{g}=3(2^{l-1}-1)\ \ , \ \ N_0^u=3(2^{l-1})\ .
\eea

We have checked these results against numerics: we have computed the
zero-point energy for all collinear configurations for $l=4,5$ and
$6$, the total number of which is 417, 7623 and 263640,
respectively. The number of ground states is given by 21, 93 and 93,
respectively; these include the $Q=0$ states. For the even case, the
odd first excited states number 24 and 96, in accordance with the
above expressions. 

What is the order parameter describing this ensemble of
configurations? In the presence of a field, the SU(2) symmetry is
broken down to U(1), but this symmetry is unbroken for the collinear
states. The ensemble, however, does break (real space) rotational
symmetry. This happens because the strings defined above cannot cross,
so that any configuration only includes dimers of at most two
different orientations. Also, shifting the dimers along one string
(e.g. by having a defect tunnel along the string) connects only dimer
configurations with the same dimer orientation absent.  The three pure
$Q=0$ (zero-string) configurations are special in this respect as they
are the only ones to contain only one dimer orientation.

The phase at the collinear point therefore is described by an order
parameter with $Z_3$ symmetry. Note that for $S=1/2$, a
$Q=\sqrt{3}\times\sqrt{3}$ ordering has been proposed,\cite{honeckerplateau}
so that there
should be a phase transition as $S$ is lowered. This appears reasonable
given the fact that at small $S$, collective quantum tunnelling around
hexagons can be expected to play a role. These processes, which are
absent in the $1/S$ expansion, will favour the
$Q=\sqrt{3}\times\sqrt{3}$ structures.\cite{msctfim}

\subsection{Coplanar states}
\begin{figure}
 \includegraphics[width=\columnwidth]{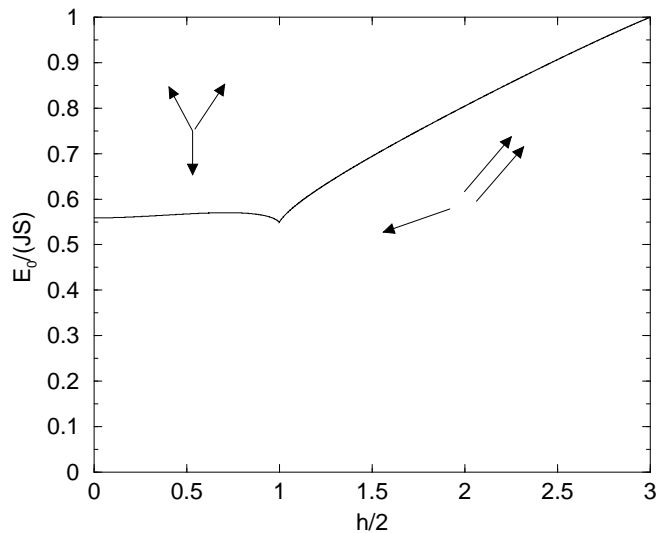} 
\caption{Zero-point energy per spin of coplanar kagome 
$Q=0$ configurations as a 
function of the magnetic field.}
\label{fig:E0Q0h}
\end{figure}

The analysis in this case is quite analogous to the preceeding
subsection. Briefly, we have checked explicitly (in intervales of $h$
of 0.2) that, among all $Q=0$ configurations, the favoured ones are
the same as found in previous studies for triangle-based lattices;
these are depicted in
Fig.~\ref{fig:E0Q0h}.\cite{chubukov,zhitokagobdo} The spin
orientations for $0\leq h\leq2$ are given as
\bea
S_1&=&(0,0,-1)\nonumber\\
S_2&=&(\sin\theta,0,\cos\theta)\\
S_3&=&(-\sin\theta,0,\cos\theta)\nonumber\\
2\cos\theta&=&h/2+1\nonumber .
\eea
For $2\leq h\leq 6$, one has 
\bea
S_1&=&(\sin\beta,0,\cos\beta)\nonumber\\
S_2=S_3&=&(\sin\alpha,0,\cos\alpha)\\ 
2\cos\alpha&=&(3+h^2/4)/h \ \ , \ 2\cos\beta=h-4\cos\alpha\nonumber .
\eea

\begin{figure}
%\rotatebox{-90}{\includegraphics[width=0.9\columnwidth]{DiffE0.ps}} 
\includegraphics[width=0.9\columnwidth]{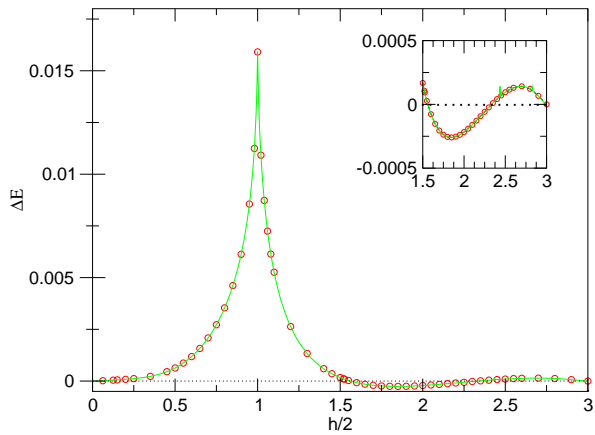} 
\caption{(colour online) 
Zero-point energy difference per spin, in units of $JS$,
between $Q=\sqrt{3}\times\sqrt{3}$ and $Q=0$ obtained from a numerical
evaluation of the Brioullin zone integral (line) of the frequencies,
and from an extrapolation of the zero-point energies of finite systems
(circles).}
\label{fig:Icepth}
\end{figure}

Fig.~\ref{fig:E0Q0h} shows the zero-point energy of the $Q=0$
configuration as a function of $h$. As for the collinear case, we
compare the energies of $Q=0$ and $Q=\sqrt{3}\times\sqrt{3}$; this is
shown in Fig.~\ref{fig:Icepth}. The data in that plot has been
obtained by numerically integrating the zero point energy using
mathematica, and by extrapolation of the finite size results.

We find that for $h<h_p$, the $Q=0$ configuration is always
energetically favoured over the $Q=\sqrt{3}\times\sqrt{3}$. For a
region of finite width for $h>h_p$, the $Q=\sqrt{3}\times\sqrt{3}$ is
in fact the one with a lower energy. In any case, note that the energy
differences are very small -- the $Q=\sqrt{3}\times\sqrt{3}$ is never
lower in energy by more than a few $10^{-3}JS$ per spin. The location
where the energy difference is appreciable is around the collinear
point, $h=2$.
 
Let us now look at the order parameter involved which exists in the
ensemble of states connected by the gauge transformations.

The vector normal to the plane of the spins defines an azimuthal angle
with respect to the field direction; this reflects the residual U(1)
symmetry. This symmetry will be broken in the coplanar states.

For $h>2$, all spins have $S^3\neq0$ and hence our dynamical symmetry
(\ref{eq:eofmznoncoll}) cannot be used to obtain a degenerate
configuration. [There do exist degeneracies between states at higher 
energies.]  As two of the three spin
directions are the same, this will again mean that only the position
of one sublattice is special, so that the order parameter will be
Z$_3\times$U(1), on top of any translational symmetry breaking.

By contrast, for $h<2$, all the three spin directions are different.
However, the spins on one sublattice points along the field direction,
and hence applying the transformation (\ref{eq:eofmznoncoll}) leads to
other degenerate configuarations.  The strings involved in the
dynamical symmetry interchange the spins with $S^3>0$ and alternating
$\pm S^1$, so that in the overall ensemble, $S_{1,2}$ have no
preferred sublattice, whereas $S_3$ does. The resulting symmetry
breaking is therefore restricted to the sublattice on which the spin
with $S^3=-1$ resides. The order parameter is again
Z$_3\times$U(1).


\subsection{Kagome-hexagon model}
\label{sec:kaghex}
In the collinear examples so far, the effective gauge Hamiltonian has
been simple in that there have existed configurations which minimise
all the terms for the shortest loop that arises. These effective
classical Hamiltonians are therefore unfrustrated. Different
scenarios, however, are possible. For instance, in the pyrochlore
slice,\cite{yao} the actual effective Hamiltonian is not dominated by
the shortest loop.

Here, we display a third way the effective Hamiltonian can turn
out. This occurs in kagome-hexagon model, in which the kagome lattice
arises from a triangular simplex lattice, that is to say, it is
considered as a lattice of corner-sharing hexagons. The zero-field
Ising case has been studied in Ref.~\onlinecite{BFG}. In the present
situation, as discussed in Sect.~\ref{sec:Effectiveh}, the collinear
manifold is selected dynamically, and the first sign of the presence
of quantum fluctuations should be incipient nematic correlations
(never long-ranged, by Mermin-Wagner) as the temperature is lowered.

For the collinear states with $L=4$, which can be mapped onto the
dimer coverings on the triangular lattice
(Fig.~\ref{fig:kagomespace}), it is the external field which
determines the direction of the collinear axis, and no spontaneous
symmetry breaking is required. 

The effective flux Hamiltonian for the resulting dimer model differs
from the previous cases in several ways. Firstly, loops which appear
in the expansion (\ref{eq:exptr}) can have odd lengths, as the simplex
lattice is non-bipartite. In fact, the shortest loop has length
three. 

Secondly, the term in the effective Hamiltonian corresponding to the
shortest loop does not act to lift the degeneracy between the dimer
states: the number of loops with an odd number of dimers equals twice
the number of dimers; all the other loops contain no dimers. This
follows from the collinear ground-state constraint (the hardcore
dimer condition) and is independent of dimer configuration.

Thirdly, whereas this first term is trivial, the next term,
corresponding to loops of length four, is frustrated: there is no
state in which all loops of length four contain an even (or all an
odd) number of dimers.  The resulting effective Hamiltonian is
therefore simple in that it is a purely classical one, but still
complicated, as it is frustrated. We leave an analysis of this model
as a subject for future work.

In closing this section, we note that, unlike in the conventional
kagome model, the selection of collinear states does {\em not} take
place classically, by the counting argument of
Ref.~\onlinecite{moecha}. Rather, this is an example where there will
be quantum, but not thermal, order by disorder.

\section{Summary and outlook}
\label{sec:summary}
In this paper, we have considered quantum fluctuations of frustrated
magnets in a magnetic field in a semiclassical framework,
focussing on the issue of dynamical symmetries. In closing, we
remark on some connections between this work and experiment, as
well as other theories of frustrated magnets.

\subsection{Energy scales and experiment}
Even though all energy scales in the semiclassical treatment
are set by $JS$, the prefactors which describe at what temperature
different effects will become visible differ greatly. Although none of
these prefactors are large, they differ considerably in their
smallness.

The tendency towards collinear ordering (Fig.~\ref{fig:eoftplat}) --
where possible -- is the largest effect. With four bonds per unit cell
having non-collinear spins, and three spins per unit cell, a
zero-point energy per spin of $0.069 JS \cos^2\theta$ implies an
effective coefficient of the biquadratic term of about $JS/20$. This
might just be visible in the most frustrated magnets. If we take a
jarosite with a spin as large as $S=5/2$, there is a factor of 50
between $JS^2$ and $JS/20$; this is comparable to the suppression of
the ordering temperature due to frustration, which is typically also
lowered by a factor of about 50 from the Curie-Weiss
temperature.\cite{willsjaro}

Selection among those collinear states is weaker -- the zero-point
energy per spin differs by {\em at most} $JS/50$. To observe this
effect, one requires not only the absence of other terms in the
Hamiltonian (beyond the classical nearest-neighbour exchange) on that
energy scale. Rather, the system might not be able to equilibrate into
the eventual ground state ensemble for reasons of `topological
glassiness',\cite{ritchey} that is to say it might not be able on
experimental timescales to tunnel through the collinear anisotropy
barriers. From the data plotted in Fig.~\ref{fig:Icepth}, it would
nonetheless seem that the best place to look for such quantum order by
disorder is near fields for which collinear configurations are
available. Luckily, a breaking of a $Z_3$ symmetry
goes along with a clear signature in the specific heat, as has been
observed elsewhere for kagome magnets with a Potts order
parameter.\cite{tsunezhito}

\subsection{(Effective) gauge theories}
The general medial lattice construction $\Lambda\rightarrow\Upsilon$
makes the emergence of gauge structures in highly frustrated magnetism
look rather natural -- the spins for systems of corner-sharing
simplices quite naturally reside on links of the simplex lattice,
$\Lambda$. A particularly striking manifestation of the tendency of
frustrated magnets to self-generate gauge descriptions is provided by
dipolar spin ice.\cite{gingrasspinice,projectiveequiv}

The classical ground-state constraint is a simple local constraint on
the sites of $\Lambda$. For the theories considered here, it is a U(1)
constraint on the lattice level, which as we have shown can be tuned
by the application of a magnetic field (the magnitude of the at each
site of $\Lambda$ is fixed by $h$). The effective long-wavelength
gauge theories depend on further details of the lattice under
consideration, such as bipartiteness.\cite{msf,subirspN} However, it
is not so surprising from the current perspective that applying a
field $h$ by itself does not change the effective theory drastically,
\cite{sviwip}
so that for example the different possible $n$-dimer models display
very similar phase structures.\cite{MStrirvb,BFG}

In the context of the gauge theories describing the correlations in
such dimer models, it is natural to think of the spins as
(gauge-invariant) electric fluxes residing on the links of
$\Lambda$. However, in the gauge-like theories of the Henley type
considered here, the collinear spins have a second, complementary role
as an Ising gauge (rather than gauge-invariant electric) field, the
magnetic fluxes $\Phi$ of which (Eq.~\ref{eq:fluxS}) determine the
zero-point energies.

It is this `dual' role of the spins in the current context which leads
to many of the unusual features of the problems under consideration.
This is yet another example of (otherwise familiar and
well-understood) effective theories in frustrated magnets exhibiting
very unusual supplementary structures.

\section*{Acknowledgements}
We are indebted to Andrey Chubukov, John Chalker, Antoine Georges,
Chris Henley, Frederic Mila, Shivaji Sondhi, Oleg Tchernyshyov and
Hirokazu Tsunetsugu for useful discussions.  RM would like to thank
Sasha Abanov, Oleg Starykh, Oleg Tchernyshyov and Hong Yao for
collaboration on related work.  This work was in part supported by the
Minist\`ere de la Recherche et des Nouvelles Technologies with an ACI
grant, and by CEFIPRA grant 2404-1.

\appendix
\section{Equation of motion in a field}
\label{app:eofmz}
A spin precesses in its effective (exchange+Zeeman) field:
\bea
\dot{S}_i&=&i[S_i,H]=-\sum_{j:i}S_j\times S_i+h\times S_i\nonumber
\eea
Summing this over $i\in\upsilon(\alpha)$ gives
\bea
\dot{L}_\alpha&=&\sum_{i\in\upsilon(\alpha)}\dot{S}_i=
-\sum_{j:i}\sum_{i\in\upsilon(\alpha)}S_j\times S_i+h\times L_\alpha \ .
\nonumber
\eea
Using $\sum_{j:i}\sab=L_\alpha+L_\beta-2\sab$, this becomes
\bea
\dot{L}_\alpha&=&-\sum_{\beta:\alpha}(L_\alpha+L_\beta)\times S_{\alpha\beta}+
h\times L_\alpha\nonumber\ .
\eea
As $\sum_{\beta:\alpha}\sab=L_\alpha$,
\bea
\dot{L}_\alpha&=&-\sum_{\beta:\alpha}
(l_\beta+\frac{h}{2})\times S_{\alpha\beta}+h\times(l_\alpha+\frac{h}{2})
\nonumber\\
&\Longrightarrow& 
(\frac{d}{dt}-\frac{1}{2} h\times)\,l_\alpha=
\sum_{\beta:\alpha}S_{\alpha\beta}\times L_\beta \nonumber\ .
\eea

As the $l$ are small deviations from the ground state, this expression
can be linearised by replacing the spins $\{S_{\alpha\beta}\}$ by
their ground state values $\{S_{\alpha\beta}^{(0)}\}$:
\bea
(\frac{d}{dt}-\frac{1}{2} h\times)\,l_\alpha=
\sum_{\beta:\alpha}S^{(0)}_{\alpha\beta}\times L_\beta  .
\eea

\section{Spectrum for $Q=0$ and 
$Q=\sqrt{3}\times\sqrt{3}$ plateau configurations}
\label{app:eofmzr3r3}
Iterating Eq.~\ref{eq:eofmz} once to eliminate the degrees of freedom
on one sublattice of triangles (corresponding to one sublattice of the
bipartite honeycomb lattice $\Lambda$) gives
\bea
(\omega-\fht)^2l^+_\alpha=
\sum_{\gamma:\beta}\sum_{\beta:\alpha}
\sab S_{\beta\gamma}l^+_\gamma\ .
\label{eq:B1}
\eea
In the following, we define the unit vectors $ e$ as $ e_1=(1,0)$,
$ e_2=(-1/2,\sqrt{3}/2)$ and $ e_2=(-1/2,-\sqrt{3}/2)$. We
have chosen twice the nearest-neighbour distance of the kagome lattice
as our unit of length.
\subsection{$Q=0$}
The resulting spectrum is 
\bea
\omega=\pm\fht\pm\sqrt{3+2(\cos{q_1}-\cos{q_2}-\cos{q_3})}\ ,
\eea
where $q_i=q^1e_i^1+q^2e_i^2$, and the two choices of $\pm$ can be made
independently. 

The zero-point energy per spin is then the average
$E_0=\langle|\omega|\rangle/2$. We find it to equal\cite{fn-confirm} 
\bea
E_0=0.5484 JS\ .
\eea

Note that the simple form $\omega=\pm \fht\pm\sqrt{}$\ is deceptive,
as the square root
can be larger than $h/2$, so that
$|\omega_1+\omega_2|=h\neq|\omega_1|+|\omega_2|$.
 
\subsection{$Q=\sqrt{3}\times\sqrt{3}$}
To compute the frequencies in this case requires solving the
characteristic polynomial of a $3\times3$ matrix. 
Denoting $\omega=\pm\fht\pm\sqrt{3+\lambda}$,
this polynomial is
given by
\bea
-\lambda^3+ f \lambda + g=0\ , \label{eq:cpoly}
\eea
Here, $f=9-2c_-$, 
$g=4+2c_3-2c_-$, with $q_i=q^1e_i^1+q^2e_i^2$ and 
\bea
c_-&=&\cos(q_1-q_2)+\cos(q_2-q_3)+\cos(q_3-q_1)\nonumber\\
c_3&=&\cos(3q_1)+\cos(3q_2)+\cos(3q_3)\nonumber\ .
\eea
The frequencies are cumbersome to write down, but a closed form
expression can be obtained by substituting the above into the formula
for solving the cubic equation Eq.~\ref{eq:cpoly}.

Finally, we find a zero-point energy of 
\bea
E_0= 0.5643 JS \ .
\eea

\end{document}